\documentclass[a4paper,twocolumn,fleqn]{article}

\usepackage{txfonts}               
\usepackage[dvipdfm]{graphicx}     
\usepackage{jpfr}                   
\usepackage{helvet}

\usepackage{amssymb}
\usepackage{setspace}
\usepackage{color}

\usepackage{ulem}

\newcommand{\iotabar}       {\mbox{$\,\iota\!\!$-}}

\newcommand {\etal}         {\textit{et al. }}
\newcommand {\mod}          {\begin{bfseries}}
\newcommand {\norm}         {\end{bfseries}}
\newcommand {\bJ}           {\mathbf{J}}
\newcommand {\bk}           {\mathbf{k}}
\newcommand {\bB}           {\mathbf{B}}
\newcommand {\bA}           {\mathbf{A}}
\newcommand {\bdS}          {\mathbf{dS}}
\newcommand {\bdl}          {\mathbf{dl}}
\newcommand {\br}           {\mathbf{r}}

\newcommand {\bI}           {\mathbf{\mathcal{I}}}
\newcommand {\bD}           {\mathbf{\mathcal{D}}}
\newcommand {\bM}           {\mathbf{\mathcal{M}}}
\newcommand {\bC}           {\mathbf{\mathcal{C}}}
\newcommand {\bF}           {\mathbf{\mathcal{F}}}

\newcommand {\bII}           {\mathbf{I}}
\newcommand {\bPI}           {\mathbf{P}}
\newcommand {\bDI}           {\mathbf{D}}
\newcommand {\bMI}           {\mathbf{M}}
\newcommand {\bCI}           {\mathbf{C}}
\newcommand {\bQI}           {\mathbf{Q}}
\newcommand {\bWI}           {\mathbf{W}}

\newcommand {\bone}         {\mathbf{1}}

\newcommand {\bsigma}       {\mathbf{\Sigma}}

\newcommand {\bppsi}        {\mathbf{p}}
\newcommand {\bfpsi}        {\mathbf{f}}
\newcommand {\grad}         {\nabla}
\newcommand {\curl}         {\nabla \times}

\usepackage[english]{babel}
\usepackage{dcolumn}  
\usepackage{bm}       

\begin{document}

\title{Model Data Fusion: developing Bayesian inversion to constrain equilibrium and mode structure}

\author{M. J. Hole, G. von Nessi, J. Bertram, J. Svensson \dag, L. C. Appel \ddag, B. D. Blackwell, R. L. Dewar and J.~Howard}

\affiliation{Research School of Physics and Engineering, Australian National University, ACT 0200, Australia. \\
\dag Max Planck Institute for Plasma Physics, Teilinstitut Greifswald, Germany. \\
\ddag Euratom/UKAEA Fusion Association, Culham Science Centre, Abingdon, Oxon OX143DB, UK.}


\email{matthew.hole@anu.edu.au}

\begin{abstract}
Recently, a new probabilistic ``data fusion'' framework based on Bayesian principles has been developed on JET and W7-AS. 
The Bayesian analysis framework folds in uncertainties and inter-dependencies in the diagnostic data and signal forward-models, 
together with prior knowledge of the state of the plasma, to yield predictions of internal magnetic structure. 
A feature of the framework, known as MINERVA (J. Svensson, A. Werner, Plasma Physics and Controlled Fusion 50, 085022, 2008), 
is the inference of magnetic flux surfaces without the use of a force balance model.  We discuss results from a new project to 
develop Bayesian inversion tools that aim to (1) distinguish between competing equilibrium theories, which capture different physics, 
using the MAST spherical tokamak; and (2) test the predictions of MHD theory, particularly mode structure, using the H-1 Heliac.   

\end{abstract}


\maketitle

\section{Introduction \label{sec:Introduction}}

With the advent of large scale neutral beam heating, magnetically confined fusion plasmas have drifted from the simple picture of ideal magnetohydrodynamics (MHD), 
which describes the plasma as a single, stationary, isotropic Maxwellian fluid. 
Due principally to neutral beam heating, several tokamak experiments now boast plasma toroidal rotation speeds that approach the thermal Mach speed 
and have significant stored energy residing in the energetic particle population produced by charge exchange with thermals.  
Motivated by these developments, a range of new descriptions have emerged that include thermal rotation \cite{Guazzotto_04} 
as well as energetic particles \cite{Hole_Dennis_09}.
Despite this, ideal MHD is still the foundation of nearly all analysis. Detailed magnetic reconstruction 
based on this treatment ignores the energetic complexity of the plasma, and can result in model-data inconsistencies,
such as thermal pressure profiles which are inconsistent with the total stored kinetic energy of the plasma.

A parallel development has been the improvement in the diversity, accuracy and resolution of plasma diagnostics.
Interpretation, however, often requires a detailed knowledge of the plasma equilibrium. For example, inference of the 
toroidal current profile $j_\phi(\psi)$ from line of sight measurements of the polarization angle requires a 
knowledge of the poloidal flux $\psi$ across the plasma. 
Formally, diagnostic forward functions relate the vector of plasma parameters $\bI$ to the measurement vector $\bD$.
For a linear system, such as toroidal current inference in a double null configuration,  $\bI$ and $\bD$ are normally related 
through a response matrix $\bM$ with additional contributions $\bC$, such that $\bD = \bM \bI + \bC$. 
Inference involves inverting this relationship to give plasma parameters $\bI$ that are consistent with the data $\bD$. A widespread technique used is 
least-square fitting, in which prior assumptions are included via a penalty term in the fit.

The confluence of higher performance plasmas with diagnostic development has however led to a dichotomy: 
data is often inconsistent with ideal MHD equilibria, sometimes misleading scientists to propose new phenomena to describe data ``artifacts''.
An example is filamentation of flux surfaces in the Rijhnuizen Tokamak, which was inferred from fluctuations in measurements of 
electron temperature \cite{Cardozo_94}.  Subsequent analysis of the Thomson scattering detection chain revealed that 
with the correct photoelectron statistics and 2D instrument profile, similar structures could arise from noise. \cite{vanMilligen_03, Cardozo_03}.

Recently, a new integrated data-modelling approach for inference of fusion plasma parameters has emerged
which offers a natural framework with which to resolve artifacts from model behaviour. 
In contrast to least square fitting, the Bayesian approach to inference involves the specification of an initial prior probability 
distribution function (pdf), $P(\bI)$, which is then updated by taking into account information that the measurements provide through 
the likelihood pdf $P(\bD|\bI)$. The result is the posterior distribution $P(\bI|\bD)$ given by Bayes' formula 
\begin{equation}
P(\bI|\bD) = P(\bD|\bI) P(\bI)/P(\bD). \label{eq:Bayes}
\end{equation}
The advantage of the Bayesian approach over traditional inversion techniques is two-fold: (i) prior knowledge, including known parameter 
inter-dependencies is made explicit, and (ii) as the formulation is probabilistic, random errors, systematic uncertainties and instrumental 
bias are integral part of the analysis rather than an afterthought.

In this work we present initial results of forward models of current tomography in the Mega Ampere Spherical Tokamak (MAST),
we formulate Bayesian inference of force balance,  and we identify plans for inference of mode structure in the H1 heliac. 
The paper is structured as follows: Sec. \ref{sec:MAST_tomography} briefly outlines the MAST experiment and 
present first results for Bayesian inference of poloidal flux and current profiles from MAST motional Stark effect (MSE) measurements.
Section \ref{sec:force_balance} outlines a Bayesian inference framework for force balance,  
synthesizes a pressure profile posterior given the flux surfaces from MSE and Thomson Scattering data, and
computes an estimate of $f(\psi)f '(\psi)$, with $f(\psi)$ the toroidal flux function. Using this, we are able to 
quantify the impact of poloidal currents on the equilibrium configuration. 
Section  \ref{sec:mode_structure} outlines a Bayesian inference model for mode structure and develops a plasma model to compute candidate
modes in the H-1 heliac. Finally, Sec. \ref{sec:conclusions} contains concluding remarks.

\section{MAST and current tomography} \label{sec:MAST_tomography}

The Mega Ampere Spherical Tokamak, which first published physics results in 2001 \cite{Sykes_01}, is one of the world's largest spherical tokamaks. 
While MAST physics and technology development have contributed across a broad range of fusion science \cite{Lloyd_07, Meyer_09}, 
two properties of MAST help motivate this research: high performance, and precision diagnostics.
It is the combination of these properties in a spherical tokamak, which has a relatively weak toroidal field and therefore 
large sensitivity to poloidal currents compared to conventional tokamaks, which helps motivate selection of MAST to
develop Bayesian inference for force balance.

Recently, both MAST neutral beam injectors have been upgraded to 3.8~MW. 
This has enabled plasma performance to be routinely lifted above  $\beta_n \approx 5$, which was reported in 2005 \cite{Hole_05}. 
MAST is equipped with an array of precision diagnostics \cite{Meyer_09}, including a high spatial resolution, single-time point, ruby Thomson Scattering system, 
a multi-time point Nd:YAG Thomson Scattering system with a coarser spatial resolution, Motional Stark effect, Charge Exchange Recombination, and 
fast magnetics \cite{Hole_Appel_Martin_09}. 

\begin{figure}[tb]
\centering
\includegraphics[width=60mm]{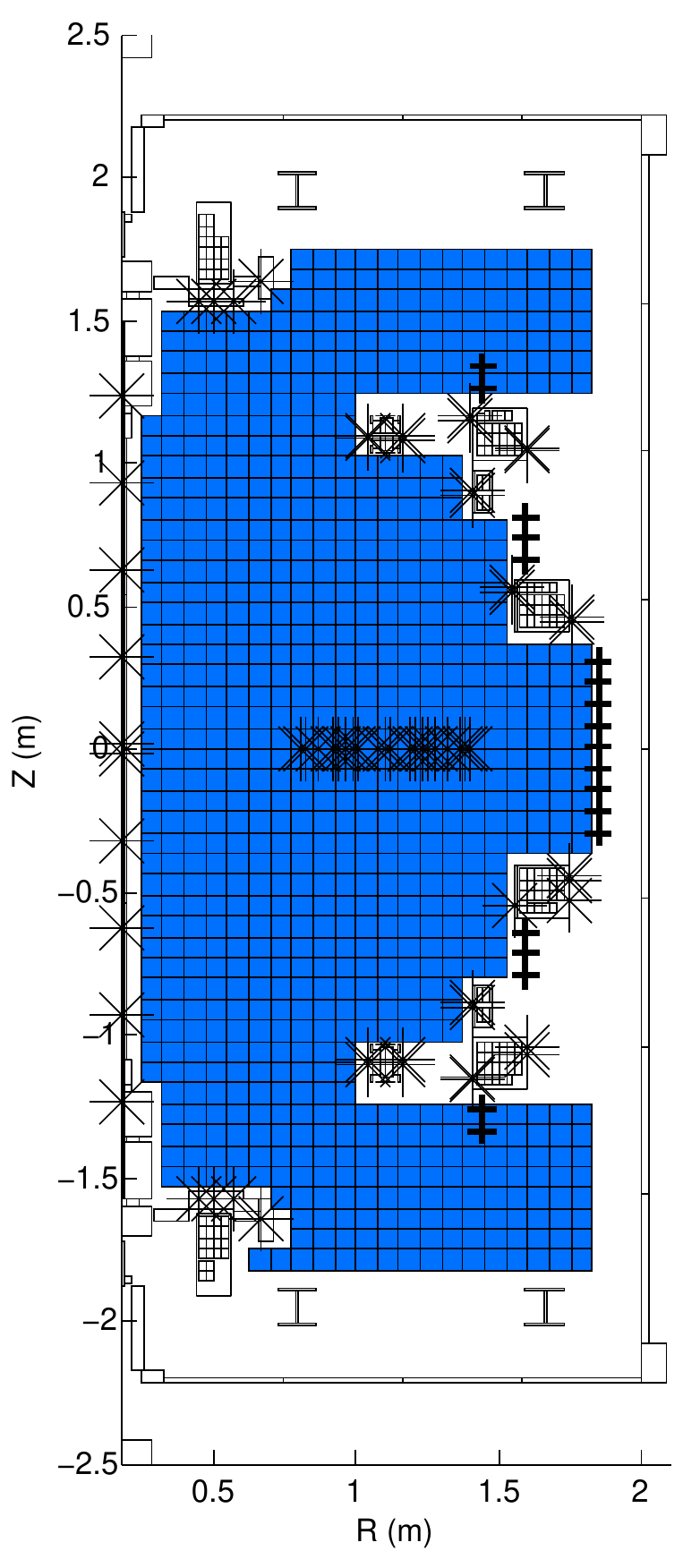}  
\caption{\label{fig:plasma_beam}: Plasma beam cross-sections fill out the MAST limiter region in the plasma beam current model.
Pick-up coils are shown as solid bars near the poloidal flux coils, flux loops as a ``*'' on both the centre column and 
poloidal field coils, and MSE as ``*'' points on a chord through the midplane.}
\end{figure}

Our development of Bayesian inference of the current profile on MAST closely follows the seminal work of Svensson and Werner \cite{Svensson:2008p1764}.
In that work, the plasma was represented as a grid of toroidal axisymmetric current beams, each with rectangular cross-section
and each beam carrying a uniform current density. In MAST, we have placed these beams so as to fill-out the entire limiter region as 
depicted in Figure \ref{fig:plasma_beam}. The magnetic field generated is a summation of Biot-Savart's law over current beams. 

A key advantage to using a series of current beams with finite cross-section to model the plasma current (as opposed to a filamentary model) 
is that the semi-analytic expressions for the corresponding magnetic field and vector potential have no singularities, even at points within 
the current beam itself. Indeed, if one were to use filaments to model the plasma current, there would be many singular points in the calculated 
magnetic field within the plasma that would make subsequent flux-surface calculations difficult and somewhat questionable \cite{Svensson:2008p1764}.

\subsection{Diagnostic Signals}
In axisymmetric devices, there exists a direct relation between the poloidal flux, $\psi$, and the toroidal component of the current vector. 
Specifically, defining the cylindrical coordinate system $(R,Z,\phi)$ with $R$ the major axis, $Z$ the vertical axis and  
$\phi$ the geometric toroidal angle, the following relation holds:
\begin{eqnarray}
\psi(R,Z)=\int \bB \cdot \bdS = \oint \bA \cdot \bdl = 2 \pi R A_\phi,\label{ds::eq1}
\end{eqnarray}
where $A_\phi$ is related to the toroidal current $j_\phi$ via Biot-Savart's law:
\begin{eqnarray}
A_\phi(\br)=\frac{\mu_0}{4\pi}\int\frac{j_\phi(\br')}{|\br-\br'|}\,dV'.\label{ds::eq2}
\end{eqnarray}
and where $\mu_0$ is the permeability of free-space.  In Eq. (\ref{ds::eq2}) $\br$ is the 
position vector $\br = R e_R + Z e_z + \phi e_\phi$, which $e_R, e_Z$, and $e_\phi$
unit vectors in the $R, Z$ and $\phi$ directions, respectively.
By inspection, the $A_R$ and $A_Z$ components due to $j_\phi$ are zero. 
Taking the curl of Eq. (\ref{ds::eq1}) and utilizing $\bB = \curl \bA$ it can be shown
\begin{eqnarray}
B_R & = & -\frac{1}{R}\frac{\partial\psi}{\partial Z},\label{ds::eq3}\\
B_Z & = & \frac{1}{R}\frac{\partial\psi}{\partial R},\label{ds::eq4}
\end{eqnarray}
where $B_R$, $B_Z$ and $B_\phi$ are the components of the magnetic field.
From Eqs. (\ref{ds::eq1})--(\ref{ds::eq4}), it can be seen that if $j_\phi$ is represented by an axisymmetric beam of rectangular cross section and 
uniform current distribution, then $\psi$, $B_R$ and $B_Z$ will vary linearly with respect to the total current going through that beam. 
The importance of this will become clear in the next two paragraphs.

The main diagnostics used to infer the current distribution are pickup coils $P_i$, full flux loops $F_i$ and the polarisation 
angle $\gamma_i$ of the emitted light from neutrally excited species during neutral beam injection due to the motional Stark effect (MSE). 
The responses of these diagnostics to a current running through a beam, $I$, is given by
\begin{eqnarray}
\hspace{-1cm} P_i(R,Z;I) & \hspace{-0.3cm}= & \hspace{-0.3cm} B_R(R,Z;I) \cos (\theta_i) + B_Z(R,Z;I) \sin(\theta_i), \label{ds::eq5} \\
\hspace{-1cm} F_i(R,Z;I) & \hspace{-0.3cm}= & \hspace{-0.3cm} \psi(R,Z;I),  \label{ds::eq6}\\
\hspace{-1cm} \tan \gamma_i(R,Z;I) & \hspace{-0.3cm}= &  \nonumber\\
 & & \hspace{-2cm} \frac{A_0 B_Z(R, Z;I) + A_1 B_R(R, Z;I) + A_2 B_\phi(R, Z;I)}{A_3 B_Z(R, Z;I) + A_4 B_R(R, Z;I) + A_5 B_\phi(R, Z;I)},  \label{ds::eq7}
\end{eqnarray}
where $\theta_i$ is the angle between a pickup coil's normal and the outboard midplane, 
and $A_0, A_1, A_2, A_3, A_4$ and $ A_5$ are constants for the particular MSE viewing geometry. 
The notation ``$;$'' denotes that the subsequent argument, in this case current $I$, is held constant. 
(By convention, the notation ``$|$'' is reserved for probability theory.) 
Both $P_i$ and $F_i$ have a linear dependence on $\psi$, $B_R$ and $B_Z$ and hence they are linearly dependent on the total current going through 
each beam of the current plasma beam model. The function $\tan\gamma_i$ has a non-linear dependence on $B_R$ and $B_Z$; however given that 
both neutral beams and MSE viewing optics all are on the mid-plane for MAST, $A_1$, $A_2$, $A_3$, $A_4$ are all approximately zero. 
The following approximation is hence used for $\tan\gamma_i$:
\begin{eqnarray}
\hspace{-1cm} \tan \gamma(R, Z;I) & = & \frac{A_0 B_Z(R, Z;I) + A_1 B_R(R, Z;I)}{A_5 B_\phi(R, Z;I)}. \label{ds::eq8}
\end{eqnarray}
where we have retained the term $A_1$.  
Since the current beam plasma model does not take into account poloidal currents, vacuum field values for $B_\phi$ are used in Eq. (\ref{ds::eq8}) 
to perform the current tomography calculations. The correction to $B_\phi$ due to poloidal currents is investigated in Sec. \ref{sec:pressure}.

Thus, given that $P_i(R,Z;I)$, $F_i(R,Z;I)$ and $\tan\gamma_i(R,Z;I)$ all have a linear dependence 
on the current flowing through $I$, one may write out a generalized prediction vector $\bPI$ for all the pickup coils, flux loops and polarization angles, as
\begin{eqnarray}
\bPI=\bMI \bII+\bCI. \label{ds::eq9}
\end{eqnarray}
Here, $\bMI$ is the response matrix of the current vector $\bII$ corresponding to all the plasma beams modelling the plasma, and 
$\bCI$ represents various other contributions to the prediction vector, which are constant relative to $\bII$.

\subsection{CAR Prior and Bayesian Inference}
In \cite{Svensson:2008p1764}, Svennson and Werner used a Conditional Auto-regressive (CAR) prior to perform Bayesian inference for 
the current beam model of the plasma, and this is the model adopted in this paper. The advantage of this choice in prior  
is that plasma beam currents are correlated to each other but in a spatially localised way. Thus, this prior has the effect of enforcing some 
smoothness between adjacent current beams, while minimising spatially long ranged effects due to manipulating the current in a particular beam. 
This is desirable in that such behaviour allows one to get a clear relation between individual current beams and diagnostic measurements. 
To construct this prior distribution, the following distribution over all current beams is considered
\begin{equation}
P(\bII) \propto \exp \left ( { -\frac{1}{2} \bII^T \bQI \bII } \right ),\label{car::eq1}
\end{equation}
where the superscript $T$ denotes the transpose, and $\bQI$ is the precision matrix. Equation (\ref{car::eq1}) is proportional to a zero-mean normal distribution, 
and so the conditional distribution of one current $I_i$ given all others $I_{-i}$ satisfies 
\begin{equation}
p(I_i|I_{-i})\propto \exp \left (  { (I_i - \sum_j \beta_{ij} I_j)^2/(2 \tau_i) } \right),\label{car::eq2}
\end{equation}
for some $\beta_{ij}$. Matching terms between Eq. (\ref{car::eq1}) and Eq. (\ref{car::eq2}) shows that $\beta_{ii}=0$, $\beta_{ij}=-Q_{ij}/Q_{ii}$ and $\tau_i=1/Q_{ii}$. 
Moreover, it turns out the symmetry of $Q$ implies that $\beta_{ij}\tau_j=\beta_{ji}\tau_i$ \cite{Besag:1974p3520}. 
We have set all variances equal. 
By setting $\beta_{ij}$ to correspond to $I_i$ having a mean that is simply the average of the currents of the beams horizontally and vertically adjacent to it, 
the precision matrix takes the form
\begin{equation}
\bQI = \frac{1}{\tau} \left ( { \bone - \frac{1}{4} \bWI } \right ),\label{car::eq3}
\end{equation}
where $\b1$ is the identity matrix and $\bWI$ the adjacency matrix with $W_{ij}=1$ if $i$ and $j$ are horizontally or vertically adjacent to current element $I_i$, 
and $W_{ij}=0$ otherwise.

From Eq. (\ref{car::eq2}), it can be seen that the CAR prior is constructed by exactly specifying the conditional distribution $p(I_i|I_{-i})$ 
and subsequently transforming this expression into a zero-mean normal distribution for $I$. CAR distributions are discussed in detail in 
\cite{Besag:1974p3520} and have the advantage that hidden parameter inter-dependencies associated with direct manipulation of the covariance 
matrix in a normal distribution are totally avoided in the CAR construction (see \cite{Svensson:2008p1764} for a discussion on this point).

The likelihood distribution function $P(\bDI|\bII)$, assuming normally distributed errors on all measurements,
is a multivariate normal distribution of the diagnostic predictions $\bPI$, given an array of diagnostic measurements $\bDI$ corresponding to the 
predictions in $\bPI$.This distribution is explicitly represented as 
\begin{equation}
\hspace{-1cm} p(\bDI|\bII) = \frac{1}{(2 \pi)^{N_d/2} |\bsigma_D|^{1/2}} \exp \left [ { -\frac{1}{2} (\bPI - \bDI)^T \bsigma^{-1} (\bPI - \bDI) } \right ]\label{car::eq3}
\end{equation}
where $N_D$ is the number of measurements, and $\bsigma$ is the covariance matrix of the measurements, which is determined experimentally. 
Finally, substituting Eq. (\ref{car::eq1}) and Eq. (\ref{car::eq3}) into Bayes' formula shows 
that the posterior distribution $p(\bII|\bDI)$ satisfies the following relation:
\begin{eqnarray}
p(\bII|\bDI) \propto \exp \left [ {-(\bPI - \bDI)^T \bsigma^{-1} (\bPI - \bDI) -\bII^T \bQI \bII} \right ].\label{car::eq4}
\end{eqnarray}
Recalling Eq. (\ref{ds::eq9}), it is known that $\bPI$ is linearly dependent on $\bII$; and thus, it can be shown that Eq. (\ref{car::eq4}) is 
proportional to a multivariate normal distribution in $\bII$ (see \cite{Svensson:2008p1764} for details). 
Hence, $p(\bII|\bDI)$ has a
simple analytic representation that be directly analysed without resorting to using Markov-chain Monte-Carlo (MCMC) algorithms. 
The speed afforded by the analytic nature of the posterior makes this analysis amenable to real-time plasma control applications.

\begin{figure}[h]
\centering
\includegraphics[width=60mm]{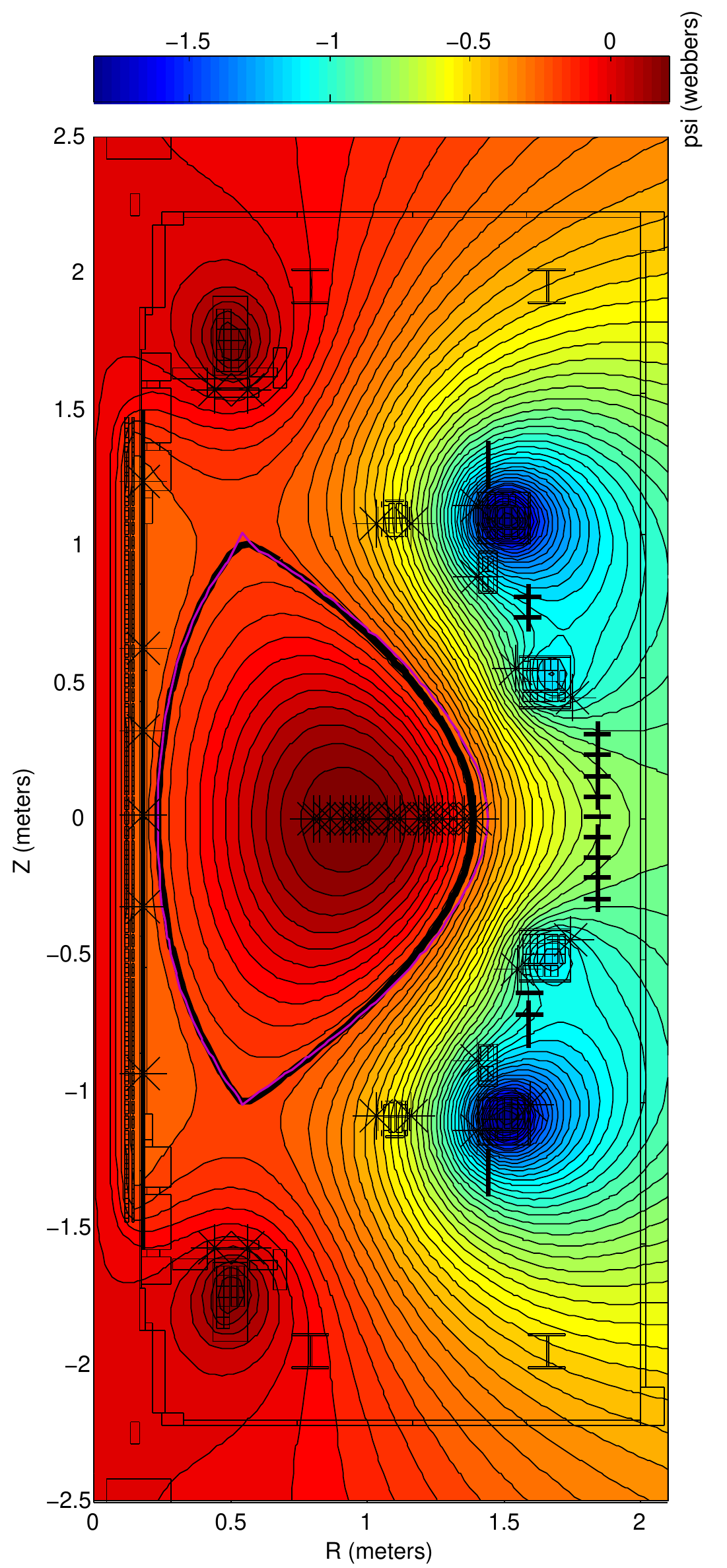}  \caption{\label{fig:psi_pol}:
Poloidal flux surfaces inferred for MAST shot \# 22254 at 320~ms using pickup coils, flux loops and MSE. }
\end{figure}

\begin{figure}[h]
\centering
\includegraphics[width=75mm]{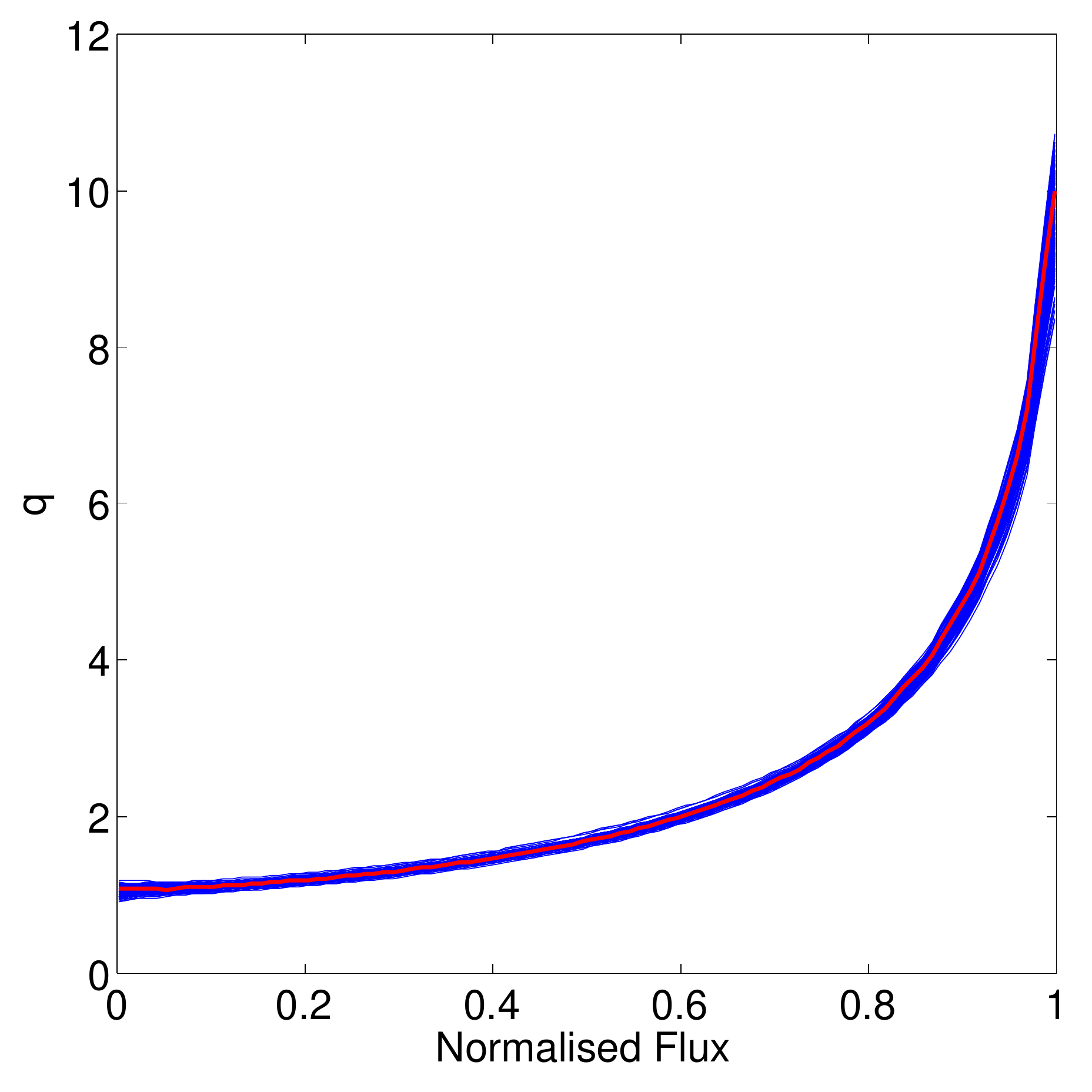}  \caption{\label{fig:qprofile}:
Safety factor, $q$ profile as a function of normalised poloidal flux found by sampling the posterior 200 times for shot \#22254 at 320ms.
The poloidal flux is normalised such that $\psi_n=0$ is the magnetic axis and $\psi_n=1$ is the edge. }
\end{figure}

With the maximum and variance of $p(\bII|\bDI)$ determined, Eqs. (\ref{ds::eq1}) and (\ref{ds::eq2}) can be used to 
construct the poloidal flux surfaces corresponding to the contours of $\psi(R,Z)$. 
Figure \ref{fig:psi_pol} shows poloidal flux surfaces from MAST discharge \# 22254 at 320~ms using pickup coils, flux loops and MSE.
Discharge \#22254 is a deuterium plasma in a double-null configuration, which was heated with 3.1~MW of neutral beam heating and a plasma current
of $I_p = 800$~kA.  The time of 320ms is analysed here as it corresponds to a 
firing of the high-resolution TS system closest to the peak $\beta$ for this shot. The figure shows a contour plot of $\psi(R,Z)$ which is 
calculated from $\bII$ corresponding to the maximum of the $p(\bII|\bDI)$ distribution. 
Overlaid on the contours are traces of the poloidal field coil cross sections and conducting surface cross sections for the MAST experiment. 
The last closed flux surface calculated from the plasma beam model is outlined in bold with the corresponding EFIT last closed flux surface overlaid in purple for comparison.
One outcome of the Bayesian approach is generation of pdfs from which the error to the fit can be inferred. 
Figure \ref{fig:qprofile} shows the inferred safety factor, $q$, profile and distribution obtained by sampling the posterior 200 times. 
The width of the fits is a characteristic measure of the error in $q$.

\section{Inference of force balance} \label{sec:force_balance}

The physics goal of our work is to exploit the improved resolution of diagnostics to
infer the validity of different force balance descriptions. Specifically, we have in mind development of a framework
that will validate energetic plasma resolved force balance models.
To build towards this goal, we commence by applying Bayesian inference to the axisymmetric Grad-Shafranov 
equilibrium description of ideal MHD force balance, $\bJ \times \bB = \grad P$.

The well known Grad-Shafranov equation \cite{Wesson_97} can be written
\begin{equation}
F(R, Z) = 0 \label{eq:force_balance}
\end{equation}
where 
\begin{eqnarray}
\hspace{-1cm} F(R,Z) & \hspace{-0.3cm} = & \hspace{-0.3cm} - \mu_0 R j_\phi(\psi) + \mu_0 R^2 p'(\psi) + \mu_0^2 f(\psi) f'(\psi), \label{eq:F} \\
\hspace{-1cm} -\mu_0 R j_\phi(\psi)  & \hspace{-0.3cm} = & \hspace{-0.3cm}  
R \frac{\partial}{\partial R} \frac{1}{R} \frac{\partial \psi}{\partial R} + \frac{\partial^2 \psi}{\partial z^2}. \label{eq:j_phi} 
\end{eqnarray}
For a real plasma, the presence of nonideal effects will mean $F(R,Z)$ is nonzero. Our long term aim is 
to compute $P(\bF|\bD)$ using Bayes formula, such that  $P(\bD|\bF) = P(\bD| \bF) P(\bF)/P(\bD)$. 
Here,the vector $\bD$ comprises all the magnetics, motional Stark effect, Thomson scattering and charge exchange recombination data, while
the column vector $\bF$ contains $F(R,Z)$ evaluated from Eq. (\ref{eq:F}) at different $(R,Z)$ across the plasma. 

An integral representation for the prior $P(\bF)$ can be obtained by using the transformation property
\begin{eqnarray}
\hspace{-1cm} P(\bF, p'(\psi), f(\psi)f'(\psi) \times 
\left | { \frac{\partial (\bF, p'(\psi), f(\psi) f'(\psi))}{\partial ( j_\phi(\psi),  p'(\psi), f(\psi) f'(\psi))}} \right |  = & & \nonumber \\
P(j_\phi(\psi), p'(\psi), f(\psi) f'(\psi)) & & 
\end{eqnarray}
and integrating both sides across $\bppsi$ and $\bfpsi$. That is,
\begin{equation}
\hspace{-1cm} 
P(\bF) = \int \frac{ P(j_\phi(\psi), p'(\psi), f(\psi) f'(\psi)) }
{ \left | { \frac{\partial (\bF, p'(\psi), f(\psi)f'(\psi))}{\partial (j_\phi(\psi),  p'(\psi), f(\psi) f'(\psi))}} \right |}
d p' d f f'(\psi) \label{eq:PF_int}
\end{equation}
where the integrand $d p' d f f'= d p'(\psi) d f(\psi) f'(\psi)$.
An ideal MHD plasma satisfies Eq. (\ref{eq:force_balance}), such that the residual force is zero, 
and so $P(\bF) = \delta(\bF)$. 
Unfortunately, the joint distribution function $P(j_\phi(\psi), p'(\psi), f(\psi) f'(\psi))$  is non-separable, preventing
direct integration of Eq. (\ref{eq:PF_int}). In future work we will fold interdependency of $j_\phi, p'$ and $f(\psi) f'(\psi)$
together to enable calculation of $P(j_\phi(\psi), p'(\psi), f(\psi) f'(\psi)$.

\subsection{Inference of toroidal flux function} \label{sec:pressure}
Some progress can be made if we assume the plasma obeys ideal force balance, and $p'(\psi)$ is assumed to be independent 
of $f(\psi) f'(\psi)$ and $j_\phi(\psi)$. In this instance, and providing we are able to 
estimate $p'(\psi)$ and its distribution from measurements, 
then $f(\psi) f'(\psi)$ across the midplane can be computed through Eq. (\ref{eq:F}). 
As a first step to inference of force balance in a real plasma, we compute $f(\psi) f'(\psi)$,
 integrate to find $f(\psi)$, and substitute this back into MINERVA. By examining the changes in the position of the magnetic axis and
$q$ profile from the recomputed solution, we are able to quantify the impact of poloidal currents on current tomography without the need for 
a separate magnetic reconstruction by EFIT. 
This folds $p'(\psi)$ into $j_\phi(\psi)$ through the influence of poloidal currents. 

For an ideal gas, the pressure is given by 
\begin{equation}
p = n_e k_B T_e + n_i k_B Ti \label{eq:p_ideal}
\end{equation}
with $n_e, n_i$ the electron and ion density, and $T_e, T_i$ the electron and ion thermal temperature. 
At $320$~ms, high resolution Thomson scattering measurements of $T_e$ and $n_e$ are available. The data
is distributed normally about each data point, and a mean and standard deviation provided.  
To obtain a pressure estimate, we have assumed that the mean values of density and temperature satisfy 
$n_i/n_e = 0.8$, and $T_i/T_e = 1.1$, as typically extracted from a charge exchange recombination measurement. We have also assumed the 
same distribution of data as Thomson Scattering. 

We have next computed a fit for the pressure $p(\psi)$ using Monte Carlo simulation \cite{NR_97}. 
The procedure is as follows: at each radial grid point, the inverse transformation method is
used to generate $n_e, T_e, n_i, T_i$ samples that satisfying the prescribed pdfs $P(n_e), P(T_e), P(n_i), P(T_i)$. 
To compute $p(\psi)$ and its distribution we have mapped each sample $p(r)$ to $p(\psi)$ using $\psi(r)$ across the inboard chord determined from MSE. 
To find a smoothed $p(\psi)$ we have then fitted a fourth order polynomial in normalized flux.
The sampling process is repeated until the pdf for $p(\psi)$ no longer changes. 

\begin{figure}[h]
\centering
\includegraphics[width=75mm]{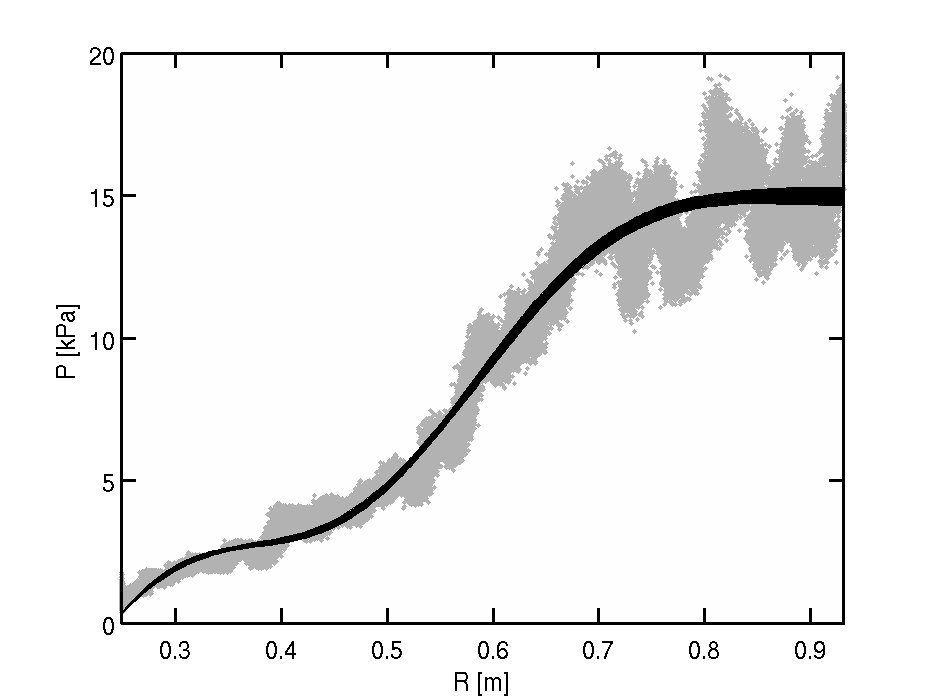}  \caption{\label{fig:pfit_inboard}: 
Inferred pressure profile $p(r)$ across inboard midplane radial chord at 320~ms in \# 22254. 
The sampled data are shown as points, and the black lines are fits to $p(\psi)$ for each sample set using a fourth order polynomial in $\psi$.  }
\end{figure}

Figure \ref{fig:pfit_inboard} shows the sampled pressure and fit for the pressure using $N_s = 2000$ samples. Overlaid
is the polynomial fit for each sample across the midplane.  The width of the polynomial fit (shown in black) yields a 
lower case estimate to the standard deviation in $p(\psi)$. 

Using the pressure profile fit of Figure \ref{fig:pfit_inboard}, 
we have sampled  $p'(\psi)$ and  $j_\phi$ across the midplane and computed $f(\psi) f'(\psi)$.
Figure \ref{fig:pfit_inboard} shows the mean and standard deviation for $p'(\psi), j_\phi$ and $f(\psi) f'(\psi)$.
For reference, the scheduler EFIT solution for this time slice is over-plotted in bold. 
While the MSE current profile is qualitatively similar to EFIT, the pressure profile is very different, particularly in the edge and core regions 
This difference is principally responsible for the difference in $f(\psi)f'(\psi)$ at the core and edge seen in panel (c).

Next, we have integrated $f(\psi) f'(\psi)$ to obtain $f(\psi)$ from panel (c). This yields a toroidal flux profile that varies approximately linearly in 
poloidal flux from $f(\psi) = -0.407$ at the edge to $f(\psi)= -0.496$ at the core. That is, poloidal plasma currents are paramagnetic, and increase 
$B_\phi$ by 22\% at the core. Qualitatively, this poloidal current should  lift the on-axis safety factor $\approx 22$\%, increase the change in 
poloidal flux across the plasma, but not change the geometry of flux surfaces or the Shafranov shift.
Work is in progress to investigate the impact of the correction to $f(\psi)$ in MINERVA.

\begin{figure}[h]
\centering
\includegraphics[width=75mm]{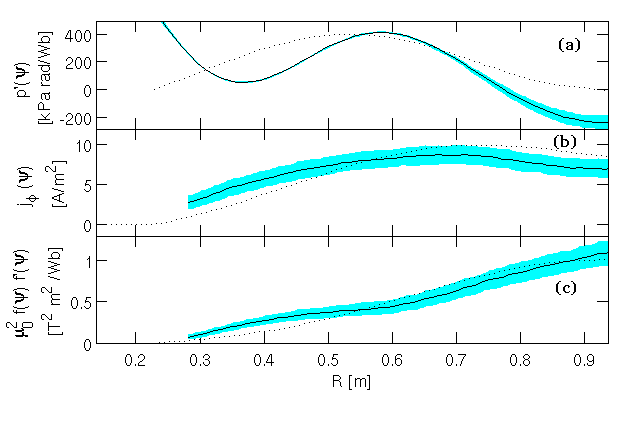}  \caption{\label{fig:ffp_inboard}: 
Inferred parameters across the midplane inboard radial chord using Monte Carlo simulation. Panel (a) shows the 
pressure gradient $p'(\psi)$, panel (b) the toroidal current and panel (c) the toroidal flux flux function $f(\psi) f'(\psi)$. }
\end{figure}

\section{Inference of mode structure} \label{sec:mode_structure}

\begin{figure}[h]
\centering
\includegraphics[width=75mm]{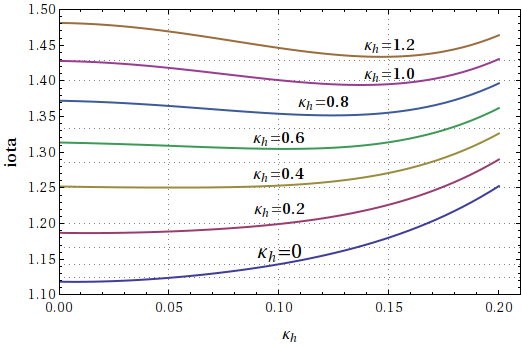}  \caption{\label{fig:iota_prof}:  
\quad Rotational transform as a function of average minor radius for the typical range of $\kappa_{h}$ in H-1NF,
 a three field-period heliac with average minor radius $r_{p}=0.2$m and major radius $R=1$m. }
\end{figure}

A second objective of the project is to test predictions of MHD mode theory, particularly mode structure. 
To achieve this goal we intend to develop a Bayesian inference tool to identify oscillations using the mode structure of a collection of candidate modes.
The low beta heliac plasma configuration of H-1NF is ideally suited to this task, as there is little uncertainty in the equilibrium, and 
precision control of the helical coil currents enables access to a wide range of magnetic rotational transform ($\iotabar$) profiles,
as shown in Fig. \ref{fig:iota_prof}. The combination thus offers a rich testing ground for Bayesian mode analysis.

For inference of mode structure, the relevant plasma parameters $\bI$ are the set of eigenmode-specifying parameters (frequency, mode numbers, radial structure moments) and 
$\bD$ is a set of Mirnov array measurements made outside the plasma. 
By solving the eigenmode problem in the plasma, and computing the vacuum-region magnetic field corresponding to a given plasma mode 
we obtain the forward model for $\bD$ given $\bI$. Given some prior $P(\bI)$, the likelihood $P(\bD | \bI)$ can then be constructed. 

In the remainder of this section we derive the forward model for Global Alfv\'{e}n Eigenmodes (GAEs) in H-1 plasmas. 
As a starting point to this sub-task, we compute GAEs in a cylindrical plasma model (coordinates $(r,\theta,z)$) incorporating a vacuum region 
in the region $r_p<r<r_w$ which is encased by a perfectly conducting wall at $r=r_{w}$. 

\begin{figure}[h]
\centering
\includegraphics[width=75mm]{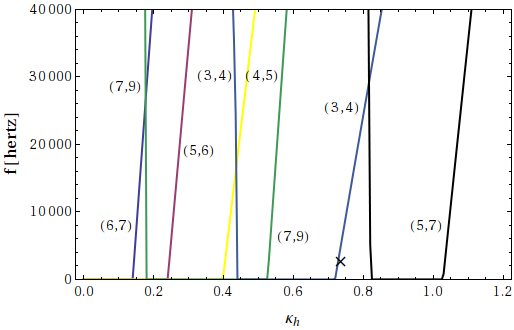}  \caption{\label{fig:mode_spectrum}: 
A plot of the minimum in the continuum $\omega_{min}/(2 \pi)$ vs $\kappa_{h}$ for a selection of mode numbers $(m,n)$. 
The cross-hairs mark the location of the eigenmode shown in Fig. \ref{fig:example_mode}.}
\end{figure}

GAEs are discrete modes that accumulate at the minimum $\omega_{min}^{2}$ of the Alfv\'{e}n continuum $\omega_{A}^{2}=k_{\parallel}^{2}v_{A}^{2}$\cite{Appert_82} 
where $k_{\parallel}$ denotes the component of the wave vector parallel to the equilibrium magnetic field and $v_{A}$ denotes the Alfv\'{e}n speed.
Expanding $k_\parallel = \bk \cdot \bB$ and substituting into $\omega_{A}^{2}=k_{\parallel}^{2}v_{A}^2$ yields  
$\omega_{A}^{2} \propto (n-\iotabar m)^2/\rho$ where $m,n$ are the poloidal and toroidal mode numbers respectively, and 
$\rho$ is the mass density profile, assumed to take the form  $\rho = \rho_0 \left ( {  1-r^2/r_{p}^2 } \right )$
with $\rho_0$ the density on-axis. 
We define $\omega_{min}^2$ to be the 
global minimum of $\omega_{A}^{2}$ over the interval $(0,r_{p})$, if such a minimum exists. That is, $\omega_{min}^2$ is the smallest 
$\omega_{A}^{2}(r_{0})$ with both ${\omega_A^2}'(r_{0})=0$ and ${\omega_{A}^{2}}''(r_{0})>0$ for some radius $r_0$ such that $0<r_{0}<r_{p}$. 
Figure \ref{fig:mode_spectrum} shows the dependency of $\omega_{min}$ on $\kappa_h$ for different $m$, $n$ modes. 
We have restricted our attention to the frequency interval $0<f<40$~kHz, where $f=\omega/(2 \pi)$, since 
we are only interested in stable modes, and current H-1 diagnostics cannot resolve frequencies above 40~kHz. 
Theoretically, for a given $\kappa_h$, GAEs lie at frequencies just below the lines in Figure \ref{fig:mode_spectrum}, accumulating at $\omega_{min}$, with an 
increasing number of radial nodes the closer the eigenfrequency lies to $\omega_{min}$.

We have modelled the plasma region by a stellarator normal-mode formulation in a cylindrical plasma \cite{Tatsuno_99,Dewar_04}, corrected for non constant
density profile. The eigenvalue equation for this problem can be written
\begin{equation}
L\varphi=\omega^{2}M\varphi, \label{eq:normal_mode}
\end{equation}
where $\omega$ is given in units of inverse Alfv\'{e}n time $\tau_A^{-1}=B_0/R\sqrt{\mu_0 \rho_0}$, with $B_0$ the field strength at the magnetic axis, and
where $\varphi$ is defined in terms of the radial element of the fluid displacement
\begin{equation}
r\xi_{r}=\exp{ \left [ {i(m\theta-\frac{n}{R}z)} \right ] }\varphi(r), \label{eq:xi_r}
\end{equation}
with $i = \sqrt{-1}$. The operators $L$ and $M$ are give by:
\begin{eqnarray}
\hspace{-1cm} L& =& -\frac{1}{r}\frac{d}{dr}(n-m\iotabar)^{2}r\frac{d}{dr}+  \nonumber \\
& & \frac{m^2}{r^2}\left[(n-m\iotabar)^2-D_{S}+\frac{\ddot{\iotabar}}{m}(n-m\iotabar)\right], \label{eq:L_op} \\
\hspace{-1cm} M& =& -\frac{1}{r}\frac{d}{dr}\frac{\rho(r)}{\rho_0} r\frac{d}{dr}+\frac{\rho(r)}{\rho_0} \frac{m^2}{r^2}. \label{eq:M_op}
\end{eqnarray}
Equation (\ref{eq:normal_mode}) is derived by averaging over helical ripple and assuming large aspect ratio. The Suydam stability parameter is given by
\begin{equation}
D_{S}=-\frac{\beta_{0}}{2\epsilon^2}p'(r)\Omega'(r) \label{eq:Suydam}
\end{equation}
where the average field line curvature is
\begin{equation}
\Omega(r)=\epsilon^{2}N \left( r^2\iotabar + 2 \int{r\iotabar} dr \right) \label{eq:field_curvature}
\end{equation}
Here $j=\sqrt{-1}$, 'prime' denotes radial derivatives, $\dot{\iotabar}=r\iotabar'$, $p(r)$ is the normalized pressure, $\epsilon$ is the inverse aspect ratio, 
$N$ is the number of turns made by the helical windings and $\beta_{0}=2\mu_{0}p_{0}/B_{0}^{2}$ is the ratio of plasma pressure to magnetic pressure at magnetic axis.

Turning now to the vacuum region, the magnetic field $B$ must satisfy $\nabla\times B=0$ and $\nabla\cdot B=0$. 
In cylindrical geometry we can write $B$ in terms of modified Bessel functions, up to a scale factor.\cite{Ross_82}  
The boundary conditions 
\begin{equation}
[[B_{r}]]=[[B_{\parallel}]]=0 \label{eq:Bjump}
\end{equation}
where $[[B]]=B_{vacuum}-B_{plasma}$, together with  
\begin{eqnarray}
\xi_{r}(0) &  = & 0,  |m|\neq1, \label{eq:xi_r0}\\ 
\xi_{r}'(0)&  = & 0,  |m|=1, \label{eq:xi_rp0}\\
B_{r}(r_{w})& = & 0, \label{eq:Br_wall}
\end{eqnarray}
relate the vacuum solution to that in the plasma. 

\begin{figure}[h]
\centering
\includegraphics[width=75mm]{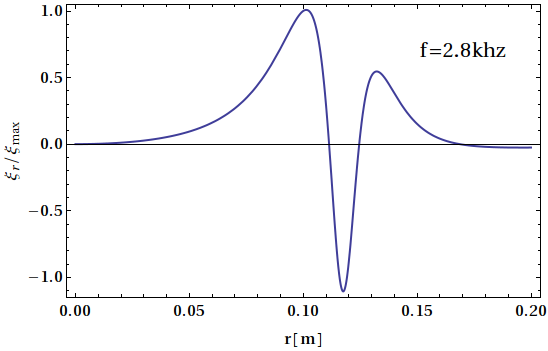}  \caption{\label{fig:example_mode}: 
The radial fluid element displacement $\xi_{r}$ as a function of minor radius for $(m,n)=(3,4)$ and $\kappa_{h}=0.73$ at the eigenfrequency indicated 
This eigenmode has angular frequency $\omega$ = 12.1 k rad s$^{-1}$}
\end{figure}

If $\omega_{min}^{2}$ appears in the plasma and is positive, a shooting method is used to find solutions to Eqs. (\ref{eq:normal_mode}),
(\ref{eq:Bjump}), (\ref{eq:Br_wall}), and Eq. (\ref{eq:xi_r0}) or Eq. (\ref{eq:xi_rp0})
with frequency $0<\omega<\omega_{min}$. More explicitly, we 
select the appropriate boundary condition for $\xi_r$ at $r=0$, and
then 'shoot' out to the plasma/vacuum interface, searching frequency space in the region $0<\omega<\omega_{min}$ for eigenfrequencies that 
are consistent with Eq. (\ref{eq:Bjump}). Figure \ref{fig:example_mode} shows an example of an $(m,n)=(3,4)$ eigenmode found in this manner, 
located at the point $(\kappa_{h},\omega)$ indicated by the cross-hairs in Figure \ref{fig:mode_spectrum}.
The global structure seen has promising features when compared to recent spectral measurements of the mode: 
the mode structure is global in radial extent,  and the radial position of the peaks nodes broadly
matches observations.

\section{Conclusions} \label{sec:conclusions}

In this work we have described a Bayesian inversion framework for inference of force-balance and mode structure. Progress has been reported in three areas:
development of Bayesian inversion for current profiles; 
inference of the toroidal flux profile assuming the plasma satisfied ideal MHD; and 
calculation of a Global Alfv\'{e}n eigenmnodes in a helical-ripple averaged cylindrical plasma model.

Based on working by Svensson \etal we have developed Bayesian inversion for current profiles and poloidal flux contours in MAST. 
The model comprises a grid of rectangular toroidally extended current beams, and is constrained by pickup coils, flux looops and MSE data.
Owing to the placement of the MSE viewing optics the tangent of the polarisation angle of emitted light is linear in $B_R$ and $B_Z$.
Signals from the pickup coils and flux loops are also linear in $B_R$ and $B_Z$. This feature means the prediction vector for the data is linear
in the currents. We have used a conditional autoregresssive prior to describe the initial current distribution, which has the 
advantage of enforcing some smoothness between adjacent current beams while minimizing spatially long ranged effects. 
Posterior calculations of the poloidal flux contours of MAST \#22254 at 320~ms illustrate good boundary agreement to EFIT.

In Sec. \ref{sec:force_balance} we outlined an inference technique for the extraction of force balance, as applied to the 
Grad Shafranov equation. Using Thomson scattering data and a Monte Carlo simulation technique, we fitted pressure profiles to the inboard
midplane chord of MAST, and computed $p'(\psi)$.  Assuming ideal MHD force balance we then calculated $f(\psi)f'(\psi)$.
While the $f(\psi) f'(\psi)$ profile is similar to EFIT, the change in $f(\psi)$ from the core to the edge is 22\%, which is slightly larger
than the EFIT value. Qualitatively, this poloidal current should  lift the on-axis safety factor by $\approx 22$\%, and increase the change in 
poloidal flux across the plasma, but not change the geometry of flux surfaces or the Shafranov shift.
At present, we are substituting the corrected $B_\phi$ profile inferred from $f(\psi)$ into MINERVA to
compute the effect of poloidal currents on the calculation of $j_\phi$ and $\psi$ surfaces. 
This constitutes a first step towards combining $j_\phi, p'(\psi)$ and $f(\psi) f'(\psi)$ to validate different equilibrium descriptions. 

Finally, we have developed an analysis approach for the inference of mode structures, and constructed a GAE model.
The model, which is in cylindrical geometry, accounts for a non constant density profile, features helical ripple averaging 
to compute the Suydam criterion and generate $\iotabar$, and includes the vacuum region.
The model has been applied
to the range of $\iotabar$ profiles accessible to H-1 plasmas, and the minimum in the continuum computed as a function of helical 
winding current ratio $\kappa_h$. The eigenfunction of a candidate $(m,n)=(3,4)$ GAE mode was determined at an angular frequency 
of 12.1~krad s$^{-1}$. The eigenfunction has similar structure to recent observations in H1, including the same number of radial 
nodes which occur at similar radial positions.  

In ongoing work we are developing forward models of Thomson scattering and charge exchange recombination spectroscopy for 
inference of force balance in MAST plasmas. We will be exploiting this information not only to improve equilibrium reconstruction, 
but to use Bayesian inference as a tool with which to resolve competing equilibrium models. In particular, we intend 
to compare force balance descriptions of ideal MHD, ideal MHD with flow \cite{Guazzotto_04}, two fluid models \cite{McClements_02},
and energetic fluid resolved equilibria \cite{Hole_Dennis_09}. The Bayesian model offers a rigorous framework with which to quantify the fit and 
thereby elucidate the important underlying physics. With regards mode structure, we intend to transform the cylindrical eigenfunction 
to a beam cross-section equilibria \cite{Davidson_95} and identify a candidate mode set which spans the different type of mode structures observed.
The next stage of this work involves deployment of forward models for mode structures into MINERVA to compute the posterior for the
mode class and parameters. On a longer time scale we envisage the cylindrical plasma model will be replaced by a 
fully 3D ideal MHD wave mode code, CAS3D \cite{Schwab_93}. As with equilibrium modelling, the importance of mode analysis
by Bayesian inversion is that it offers a rigorous framework in which to identify wave mode structures, some of which 
have deleterious effects on plasma performance.  

\section*{Acknowledgments}
\noindent This work was jointly funded by the Australian Government through International Science Linkages Grant CG130047,  the
Australian National University, the United Kingdom Engineering and Physical Sciences Research Council, and by the
European Communities under the contract of Association between EURATOM and UKAEA.
The views and opinions expressed herein do not necessarily reflect those of the European Commission.

\bibliographystyle{prsty}


\begin{thebibliography}{10}

\bibitem{Guazzotto_04}
L. Guazzotto, R. Betti, J. Manickam, and S. Kaye, Phys. Plasmas {\bf 11},  604
  (2004).

\bibitem{Hole_Dennis_09}
M.~J. Hole and G. Dennis, Plas. Phys. Con. Fus. {\bf 51},  035014  (2009).

\bibitem{Cardozo_94}
N.~J.~L. Cardozo {\it et~al.}, Phys. Rev. Lett. {\bf 73},  256  (1994).

\bibitem{vanMilligen_03}
B.~P. van Milligen, I.~G.~J. Classen, and C.~J. Barth, Rev. Sci. Instrum. {\bf
  74},  3998  (2003).

\bibitem{Cardozo_03}
N.~J.~L. Cardozo {\it et~al.}, Phys. Rev. Lett. {\bf 90},  0031  (2003).

\bibitem{Sykes_01}
A. Sykes {\it et~al.}, Phys. Plas. {\bf 8},  2101  (2001).

\bibitem{Lloyd_07}
B. Lloyd and \textit{et al}, Nuc. Fus. {\bf 47},  S658  (2007).

\bibitem{Meyer_09}
H. Meyer and \textit{et al}, Nuc. Fus. {\bf 49},  104017  (2009).

\bibitem{Hole_05}
M. Hole {\it et~al.}, Plas.Phys. Con. Fus. {\bf 47},  581  (2005).

\bibitem{Hole_Appel_Martin_09}
M.~J. Hole, L.~C. Appel, and R. Martin, Rev. Sci. Instrum.  (2009), submitted.

\bibitem{Svensson:2008p1764}
J. Svensson and A. Werner, Plas. Phys. Cont. Fus. {\bf 50},  085002  (2008).

\bibitem{Besag:1974p3520}
J. Besag, Journal of the Royal Statistical Society. Series B {\bf 36},  192
  (1974).

\bibitem{Wesson_97}
J. Wesson, {\em Tokamaks}, 2nd ed. (Oxford Univ. Press, Oxford, 1997).

\bibitem{NR_97}
W.~H. Press, S.~A. Teukolsky, W.~T. Vetterling, and B.~P. Flannery, {\em
  {Numerical Recipes in Fortran 77: The Art of Scientific Computing}}, 2nd ed.
  (Univ. of Cambridge Press, Cambridge, 1997).

\bibitem{Appert_82}
K. Appert, R. Gruber, F. Troyon, and J. Vaclavik, Plas. Phys. {\bf 24},  1147
  (1982).

\bibitem{Tatsuno_99}
T. Tatsuno and M. Wakatani, Nucl. Fusion {\bf 39},  1391  (1999).

\bibitem{Dewar_04}
R.~L. Dewar {\it et~al.}, Phys. Rev. E {\bf 70},  0066409  (2004).

\bibitem{Ross_82}
D.~W. Ross, G.~L. Chen, and S.~M. Mahajan, Phys. Fluids {\bf 25},  652  (1982).

\bibitem{McClements_02}
K.~G. McClements and A. Thyagaraja, Mon. Not. R. Astron. Soc. {\bf 323},  733
  (2001).

\bibitem{Davidson_95}
M.~G. Davidson, R.~L. Dewar, H.~J. Gardner, and J. Howard, Aus. J. Phys. {\bf
  48},  871  (1995).

\bibitem{Schwab_93}
C. Schwab, Phys. Fluids B - Plas. Phys. {\bf 9},  3195  (1993).

\end{thebibliography}

\end{document}